\documentclass{article}
\usepackage[utf8]{inputenc}
\usepackage[a4paper]{geometry}
\usepackage{float}
\usepackage{booktabs}
\usepackage{amsmath}
\usepackage{amssymb}
\usepackage{graphicx}
\usepackage{subscript}

\makeatletter

\providecommand{\tabularnewline}{\\}

\newcommand{\lyxaddress}[1]{
\par {\raggedright #1
\vspace{1.4em}
\noindent\par}
}

\usepackage{setspace}
\usepackage{etoolbox}
\AtBeginEnvironment{align}{\setcounter{subeqn}{0}}
\newcounter{subeqn} %

\usepackage[small,bf]{caption}\usepackage{color}\usepackage[usenames,dvipsnames]{xcolor}
\usepackage{booktabs}
\usepackage{array}
\usepackage{paralist}
\usepackage[bottom]{footmisc}
\usepackage{multicol}\usepackage{todo}
\usepackage[version=3]{mhchem}
\usepackage{cite}

\makeatother

\begin{document}

\title{Accommodation of Tin in Tetragonal ZrO\textsubscript{2}}

\author{B. D. C. Bell\textsuperscript{a}, S. T. Murphy\textsuperscript{b},
P. A. Burr\textsuperscript{a, c}, R. W. Grimes\textsuperscript{a},
M. R. Wenman{*} \textsuperscript{a}}

\maketitle

\lyxaddress{\textsuperscript{a}Department of Materials and Centre for Nuclear
Engineering, Imperial College, London, SW7 2AZ, UK}

\lyxaddress{\textsuperscript{b}Department of Physics and Astronomy, University
College London, Gower Street, London, WC1E 6BT, UK}

\lyxaddress{\textsuperscript{c}Institute of Materials Engineering, Australian
Nuclear Science \& Technology Organisation, Menai, New South Wales
2234, Australia}
\begin{abstract}
Atomic scale computer simulations using density functional theory
were used to investigate the behaviour of tin in the tetragonal phase
oxide layer on Zr-based alloys. The $\mbox{\ce{S\ensuremath{n_{Zr}^{\times}}}}$
defect was shown to be dominant across most oxygen partial pressures,
with $\mbox{\ce{S\ensuremath{n_{Zr}^{''}}}}$ charge compensated by
$\mbox{\ce{\ensuremath{V_{O}^{\bullet\bullet}}}}$ occurring at partial
pressures below $10^{-31}$~atm. Insertion of additional positive
charge into the system was shown to significantly increase the critical
partial pressure at which $\mbox{\ce{S\ensuremath{n_{Zr}^{''}}}}$
is stable. Recently developed low-Sn nuclear fuel cladding alloys
have demonstrated an improved corrosion resistance and a delayed transition
compared to Sn-containing alloys such as Zircaloy-4. The interaction
between the positive charge and the tin defect is discussed in the
context of alloying additions such as niobium and their influence
on corrosion of cladding alloys.
\end{abstract}
\begin{doublespace}

\section{Introduction}

\subsection{Corrosion of zirconium}

Zirconium alloys are used in pressurised water reactors due to their
low capture cross-section for thermal neutrons, high corrosion resistance
and acceptable mechanical properties. Initially, corrosion follows
an approximately cubic rate law \cite{Hillner1977}. After a few microns
of oxide growth, there is a sudden increase in the corrosion rate
as the protective layer breaks down; this process is generally referred
to as transition, or break-away corrosion. Post-transition, a reduced
corrosion rate is again observed. This process repeats in a cycle;
the time from initial corrosion to transition and subsequent later
transitions is also composition dependent \cite{Hillner1977,Bryner1979}.

During corrosion hydrogen is produced, which arrives at the metal
oxide interface. Hydrogen has a low solubility in ZrO\textsubscript{2}
so that after recombining with electrons, the hydrogen atoms move
either into the coolant, or into the oxygen-saturated $\alpha$-Zr
beneath the oxide (in which hydrogen exhibits a much higher solid
solubility \cite{Miyake1999}). When the solubility limit for hydrogen
in $\alpha$-Zr is exceeded, zirconium hydrides precipitate. The fraction
of hydrogen produced during corrosion, that moves into the cladding
metal, is called the hydrogen pick-up fraction (HPUF) and is alloy
dependent. It has been suggested \cite{Couet2011} that alloys which
exhibit a lower HPUF have a more electrically conductive oxide layer,
allowing ion-electron recombination to occur further from the metal-oxide
interface thereby lowering the probability of the hydrogen entering
the cladding metal. A lower HPUF is desirable, since there is a regulatory
upper limit to the amount of hydrogen allowed in the cladding metal
\cite{Raynaud2011}.

\subsection{Oxide layer composition}

Various experimental techniques have demonstrated that the pre-transition
oxide layer formed on the surface of zirconium alloys is composed
of several distinct regions. From the outside inwards: an outer layer
of predominantly columnar monoclinic ZrO\textsubscript{2}, a 50-80~nm
layer containing equiaxed tetragonal ZrO\textsubscript{2}, often
a 100-200~nm layer of sub-stoichiometric Zr-oxide and finally $\alpha$-Zr
metal (saturated with oxygen close to the metal-oxide interface) \cite{Beie1993,Preuss2011,Garner2014}.

Early studies of the sub-stoichiometric oxide layer were unable to
identify the crystal structure other than it resembled distorted $\alpha$-Zr
\cite{Anada1996}. Subsequent TEM work hinted at several possible
sub-stoichiometric zirconia structures, such as cubic ZrO \cite{Moseley1981,Furuta1980}
or an ordered oxygen solid solution within $\omega$-Zr with an approximate
stoichiometry of Zr\textsubscript{3}O \cite{Iltis1991}. Random structure
searching of the Zr-system using density functional theory have identified
previously unknown energetically stable structures \cite{Lumley2013,Wang2013},
recent studies have identified a metastable ZrO structure, with an
HCP lattice similar to $\alpha$-Zr \cite{Puchala2013,Nicholls2014}.

It is generally agreed that an increased proportion of tetragonal
ZrO\textsubscript{2} phase is found adjacent to the metal-oxide interface,
where it is stabilised by a combination of grain size, compressive
stress and the presence of alloying (i.e. dopant) elements \cite{Ding1993,Petigny2000,Yilmazbayhan2004}.
Immediately post-transition, oxides have been shown to contain very
little tetragonal or sub-stoichiometric zirconia away from the metal
oxide interface, suggesting that these phases are transformed to monoclinic
ZrO\textsubscript{2} during transition \cite{Anada1996}.

At the metal oxide interface, the oxygen partial pressure is expected
to be extremely low, however the limited amount of atom probe \cite{Hudson2009}
and TEM \cite{Ni2012} studies thus far have demonstrated that a significant
variation from ZrO\textsubscript{2} stoichiometry is not observed;
instead a sharp transition occurs from ZrO\textsubscript{2} to ZrO
over a distance of a few nm, with a more gradual change from ZrO to
oxygen-saturated $\alpha$-Zr occurring over a distance of tens of
nm.

\subsection{Oxide phase stability\label{sub:Oxide-phase-stability}}

Various investigations have demonstrated the presence of high compressive
stresses (0.2-3~GPa) in the oxide layer in the plane of the oxide-metal
interface~\cite{Roy1970a,Polatidis2013,Antonia1991,Garzarolli1991},
due to the Pilling-Bedworth ratio of 1.56 on converting from zirconium
to ZrO\textsubscript{2}. Experimental work had shown that applied
hydrostatic stress can stabilise the metastable tetragonal ZrO\textsubscript{2}
phase due to it having a smaller volume than the monoclinic phase
\cite{Bouvier2002,Block1985}. Recent synchrotron x-ray diffraction
work performed by Ortner~\textit{et~al.} \cite{Ortner2014} has
demonstrated a significant variation in the stress through the oxide
layer, with higher in-plane compressive stresses observed close to
the metal-oxide interface. Similar work performed by Polatidis~\textit{et~al.}
also identified a discontinuity in the stress profile, with a sudden
increase in compressive stress at a depth within the oxide layer,
which had previously been identified, using electron microscopy, as
the approximate point of transition from monoclinic to tetragonal
rich ZrO\textsubscript{2} \cite{Polatidis2013}.

Small grain sizes (typically less than 30 nm) can also result in tetragonal
phase stability with no additional environmental factors \cite{Ding1993,Aldebert1985,Barberis2001}.
Qin\textit{~et~al.}~\cite{Qin2007} combined these theories into
a thermodynamic model, suggesting that in any given environment there
is a critical grain size below which the tetragonal phase is stable.
Applied stress increases this critical grain size allowing more tetragonal
phase to be stabilised.

Dopant stabilisation can occur due to the incorporation of larger
cations that expand the oxide lattice, or by the incorporation of
lower valence cations, which stabilise the tetragonal and cubic fluorite
phases by the incorporation of oxygen vacancies. While the concentrations
of dopant elements required for complete cubic phase stabilisation
(e.g.~20~mol\%~yttrium~\cite{Bogicevic2001}) are far higher than
the concentration of alloying elements in any common zirconium based
alloys, concentrations as low as 2 mol\%~yttrium can stabilise the
tetragonal phase \cite{Xia2010}. When the other complementary stabilisation
methods (stress and grain size) are present, it is possible that even
1-2\% of trivalent and divalent cations could have a stabilising effect.

\subsection{Tin as an alloying addition}

Tin was added initially to reduce the detrimental effects of nitrogen
and carbon impurities, and is still present in Zircaloy-2 and Zircaloy-4
as it also improves strength and creep resistance \cite{Hong1996}.
It has, however, been demonstrated that the removal of tin from modern
cladding alloys can increase the time until transition occurs and
lower HPUF \cite{Wei2013,Schefold2003,Takeda2000,Barberis1999}.

Unlike many other alloying elements, the amount of tin included is
lower than the solid solubility limit in $\alpha$-Zr, so that a homogeneous
distribution throughout the cladding matrix is expected. Wei~\textit{et~al.}~\cite{Wei2013}
observed that in the oxide layers of various Zr-Sn-Nb alloys, those
with lower tin content exhibited a smaller proportion of pre-transition
tetragonal phase. A reduction in the corrosion rate, due to a delayed
transition, was also observed, implying a correspondence with tetragonal
phase fraction, tin content and corrosion rate, which had not previously
been suggested.

In this study, the behaviour of tin as a substitutional and interstitial
defect in tetragonal ZrO\textsubscript{2} is investigated using computer
simulation. Quantum mechanical calculations using density functional
theory are performed and comparisons are made to previous experimental
data concerning the corrosion of tin-containing zirconium alloys.

\section{Methodology}

Simulations were performed using the density functional theory based
CASTEP~6.11 code \cite{Clark2005}. Ultra-soft pseudo potentials
with a cut-off energy of 550~eV were used throughout. The Perdew,
Burke and Ernzerhof \cite{Perdew1996} parametrisation of the generalised
gradient approximation was employed to describe the exchange correlation
function. A Monkhorst-Pack sampling scheme \cite{Monkhorst1976} was
used for the integration of the Brillouin Zone, with a minimum k-point
separation of 0.045~Å\textsuperscript{-1}. The simulations employed
density mixing using the Pulay method \cite{Pulay1980}.

The energy convergence criterion for self-consistent calculations
was set to $1\times10^{-8}$~eV and the maximum allowed forces between
ions was $1\times10^{-2}$~eV/Å. All simulations were performed until
a maximum difference in energy of $1\times10^{-5}$~eV and atomic
displacement of $5\times10^{-4}$~Å between iterations was achieved.

Non-defective structures were relaxed under constant pressure to the
above convergence criteria. All defective structures were generated
from pre-relaxed non-defective structures, and were energy minimised
under constant volume (cell parameters constrained to maintain the
shape and volume of the perfect supercell) in order to approximate
dilute conditions. A supercell was formed from $3\times3\times2$
repetitions of the tetragonal ZrO\textsubscript{2} unit cell in the
x, y and z directions respectively. This resulted in a 108~atom supercell,
which offered a reasonable compromise between reducing finite size
effects and computation time.

\subsection{Defect formation energies}

The defect formation energies ($E^{\text{{f}}}$) of intrinsic defects
were calculated using Equation~\ref{eq:formation_intrinsic}, where
a point defect $X$ with charge $q$ is formed in a perfect cell.

\begin{equation}
E_{\text{{\ensuremath{X^{q}}}}}^{\text{{f}}}=E_{\text{{\ensuremath{X_{Zr}^{q}}}}}^{\text{{DFT}}}-E_{\text{{perfect}}}^{\text{{DFT}}}\pm\underset{i}{\sum}n_{i}\mu_{i}+q(E_{\text{VBM}}+\mu_{e})+E_{\text{MP}}\label{eq:formation_intrinsic}
\end{equation}
$E_{\text{{perfect}}}^{\text{{DFT}}}$ is the energy of the perfect
cell, $E_{\text{{\ensuremath{X_{Zr}^{q}}}}}^{\text{{DFT}}}$ is the
energy of the defective cell, $E_{\text{VBM}}$ is the valence-band
maximum (VBM) of the perfect supercell, $n_{i}$ is the number of
atoms added/removed, $\mu_{i}$ is the chemical potential of the defect
species added/removed. $E_{\text{MP}}$ is an energy correction calculated
using the Makov-Payne method \cite{Makov1995} to account for the
electrostatic self-interaction of defects caused by the use of periodic
boundary conditions and a finite supercell size. $\mu_{e}$ is the
chemical potential of electrons relative to $E_{\text{VBM}}$; taking
values of $\mu_{e}$ at the VBM and conduction-band minimum, the defect
formation energy can be plotted as a function of $\mu_{e}$ across
the band gap of the material \cite{Tahini2013,Murphy2014}.

\subsection{Chemical potentials}

For a given set of conditions, the sum of the chemical potentials
($\mu$) per formula unit of the constituent species equals the total
Gibbs free energy of the solid ZrO\textsubscript{2}, leading to the
following relationship, 
\begin{equation}
\mu_{\text{{Zr\ensuremath{O_{2}}(s)}}}=\mu_{\text{{Zr}}}(T,p_{\text{{\ensuremath{O_{2}}}}})+\mu_{\text{{\ensuremath{O_{2}}}}}(T,p_{\text{{\ensuremath{O_{2}}}}})
\end{equation}
Where $T$ is the temperature and $p_{\text{{\ensuremath{O_{2}}}}}$
is the oxygen partial pressure. In order to avoid the well-known difficulties
regarding the inability of DFT to accurately describe the O\textsubscript{2}
molecule, we adopt an approach first used by Finnis~et~al.~\cite{Finnis2005}.
This avoids the necessity to calculate the oxygen chemical potential
in DFT by instead using the known experimental formation energy of
the ZrO\textsubscript{2} oxide, and calculating the chemical potential
of oxygen at standard temperature and pressure using the following
relationship:

\begin{equation}
\Delta G_{\text{{f}}}^{\text{{Zr\ensuremath{O_{2}}}}}=\mu_{\text{{Zr\ensuremath{O_{2}}(s)}}}-\mu_{\text{{Zr(s)}}}-\mu_{\text{{\ensuremath{O_{2}}}}}(T^{\circ},p_{\text{{\ensuremath{O_{2}}}}}^{\circ})
\end{equation}
We assume the Gibbs free energy is independent of temperature for
the solid species, however for the gaseous O\textsubscript{2}, this
assumption does not hold true. Instead the oxygen chemical potential
at the temperature of interest $\mu_{\text{{\ensuremath{O_{2}}}}}(T,p_{\text{{\ensuremath{O_{2}}}}})$
is extrapolated from the standard state $\mu_{\text{{\ensuremath{O_{2}}}}}(T^{\circ},p_{\text{{\ensuremath{O_{2}}}}}^{\circ})$
using the ideal gas relationship. The full expression for the oxygen
chemical potential is as follows:

\begin{equation}
\mu_{\text{{\ensuremath{O_{2}}}}}(T,p_{\text{{\ensuremath{O_{2}}}}})=\mu_{\text{{\ensuremath{O_{2}}}}}(T^{\circ},p_{\text{{\ensuremath{O_{2}}}}}^{\circ})+\Delta\mu(T)+\frac{1}{2}k_{B}\log\left(\frac{p_{\text{{\ensuremath{O_{2}}}}}}{p_{\text{{\ensuremath{O_{2}}}}}^{\circ}}\right)
\end{equation}
and the rigid-dumbell ideal gas for $\Delta\mu(T)$ can be given by:

\begin{equation}
\Delta\mu(T)=-\frac{1}{2}\left(S_{\text{{\ensuremath{O_{2}}}}}^{\circ}-C_{\text{{P}}}^{\circ}\right)\left(T-T^{\circ}\right)+C_{\text{{P}}}^{\circ}T\log\left(\frac{T}{T^{\circ}}\right)
\end{equation}
Where $S_{\text{{\ensuremath{O_{2}}}}}^{\circ}$ is the molecular
entropy at standard temperature and pressure and $C_{\text{{P}}}^{\circ}$
is the constant pressure heat capacity of oxygen gas. Values for these
two properties were obtained from the literature with $S_{\text{{\ensuremath{O_{2}}}}}^{\circ}=0.0021$~eV/K
and $C_{\text{{P}}}^{\circ}=7k_{B}=0.000302$~eV/K~\cite{Weast1984}.

Tetragonal ZrO\textsubscript{2} is not stable under standard conditions,
so the free energy of formation was calculated by adding the energy
difference between monoclinic and tetragonal ZrO\textsubscript{2}
(as calculated at 0~K by DFT) to the experimentally determined value
for the free energy of formation of monoclinic ZrO\textsubscript{2},
this resulted in a value of $-11.41$~eV.

\subsection{Brouwer Diagram\label{sub:Brouwer-Diagram}}

The sum of all defects each multiplied by their charge must equal
zero, since there is no overall charge on the crystal. This can be
expressed as follows:

\begin{equation}
\underset{i}{\sum}q_{i}c_{i}-N_{c}\exp\left(-\frac{E_{g}-\mu_{e}}{k_{B}T}\right)+N_{v}\exp\left(-\frac{\mu_{e}}{k_{B}T}\right)=0\label{eq:charge_neutrality}
\end{equation}
where the first term is the sum of the charges of all ionic defects,
the second term is the electron concentration and the third term the
hole concentration in the crystal. $N_{c}$ and $N_{v}$ are the density
of states for the conduction and valence bands and $\mbox{\ensuremath{E_{g}}}$
is the band gap of the crystal. Tetragonal ZrO\textsubscript{2} is
an insulating material and so the concentrations of electrons and
holes are expected to be sufficiently low that Boltzmann statistics
are appropriate. The formation energy for an electron in the conduction
band will be slightly lower than the value calculated by $E_{g}-\mu_{e}$
due to self trapping of electrons. However, in a wide band gap insulator
such as ZrO\textsubscript{2} the difference in energy will be minimal
and so this is an acceptable approximation.

Using this relationship, the electron chemical potential that ensures
charge neutrality in the system can be calculated for any given elemental
chemical potential. Using this, the concentration of individual defects
can be calculated and by plotting these as a function of oxygen partial
pressure, a Brouwer diagram is constructed.

\section{Results and Discussion}

\subsection{Interstitials}

Potential interstitial sites were chosen by considering the Wyckoff
positions of the space group for tetragonal ZrO\textsubscript{2}
($P4_{2}/nmc$). The 2b and 4c positions are very similar, however
both were included to investigate whether there was a preference to
be closer to a zirconium cation (as in the case of the 4c position)
or a position equidistant between two (as in the 2b position). An
interstitial ion was placed in each non-identical site and an overall
charge was applied to the supercell corresponding to the charge state
of the interstitial ion; -2 to 0 for oxygen, 0 to +4 for zirconium
and tin. A list of the non-identical sites considered, along with
the corresponding lattice co-ordinates is reported in Table~\ref{tab:interstitial-sites}.

A successful convergence was only achieved for oxygen in the 2b location,
suggesting that the other sites are not stable for any charge state.
Convergence was achieved for zirconium and tin interstitials on all
sites; for both the 2b site was preferred. Tin and zirconium interstitials
placed at the 4c site migrated to the 2b site during the simulation
and the 4d site was not energetically favourable for any interstitials,
generally exhibiting a formation energy several eV higher than the
2b site.

\begin{table}[h]
\caption{A list of the non-identical Wyckoff positions for tetragonal ZrO\textsubscript{2}
considered as possible interstitial sites.\label{tab:interstitial-sites}}

\hfill{}%
\begin{tabular}{cccc}
\toprule 
Position & \textit{a} & \textit{b} & \textit{c}\tabularnewline
\midrule
\midrule 
2b & 0 & 0 & 0.5\tabularnewline
\midrule 
4c & 0 & 0 & -0.25\tabularnewline
\midrule 
4d & 0.5 & 0 & 0\tabularnewline
\bottomrule
\end{tabular}\hfill{}
\end{table}

\subsection{Defect formation energies}

\begin{figure}
\hfill{}\includegraphics[width=0.5\textwidth]{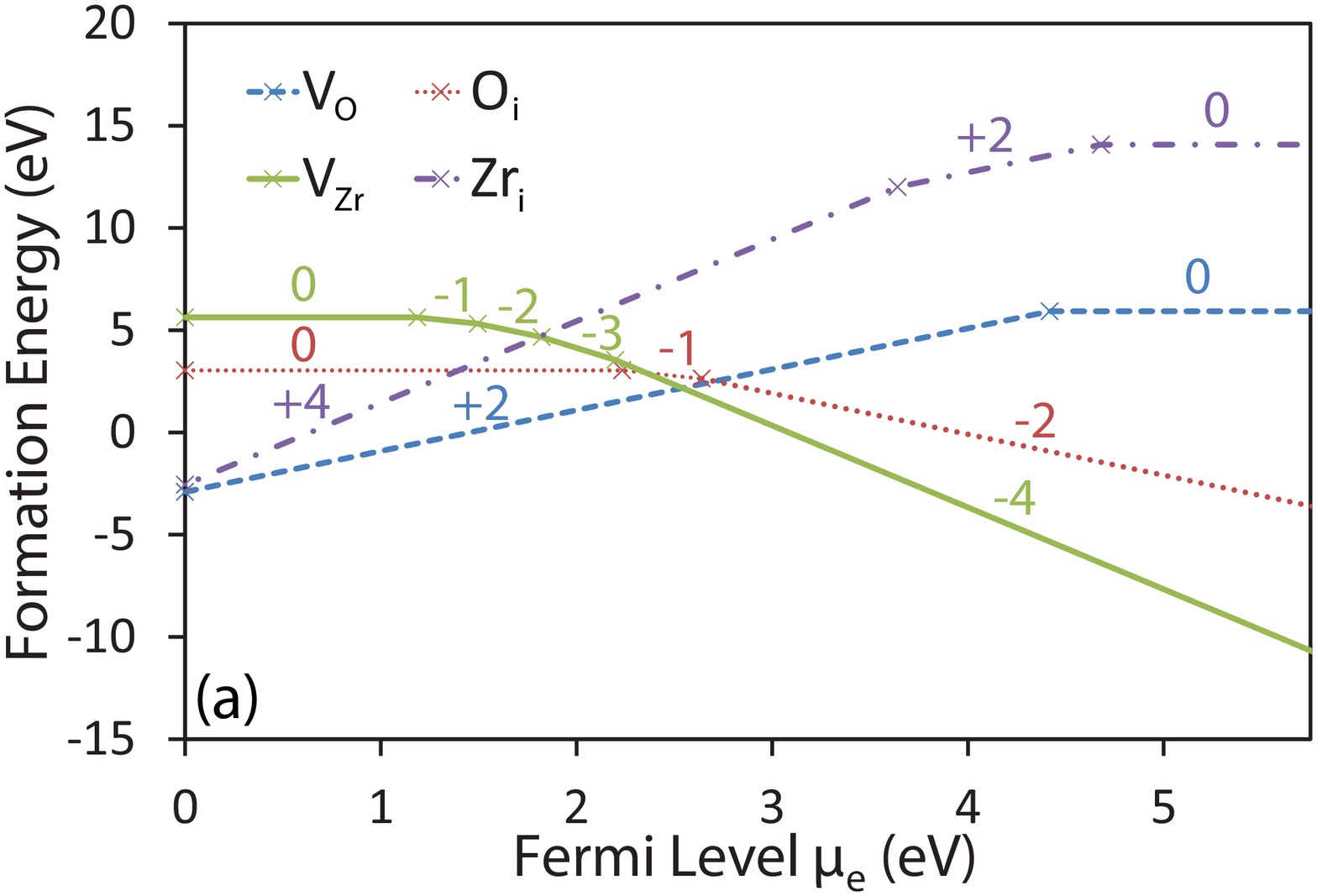}\hfill{}\includegraphics[width=0.5\textwidth]{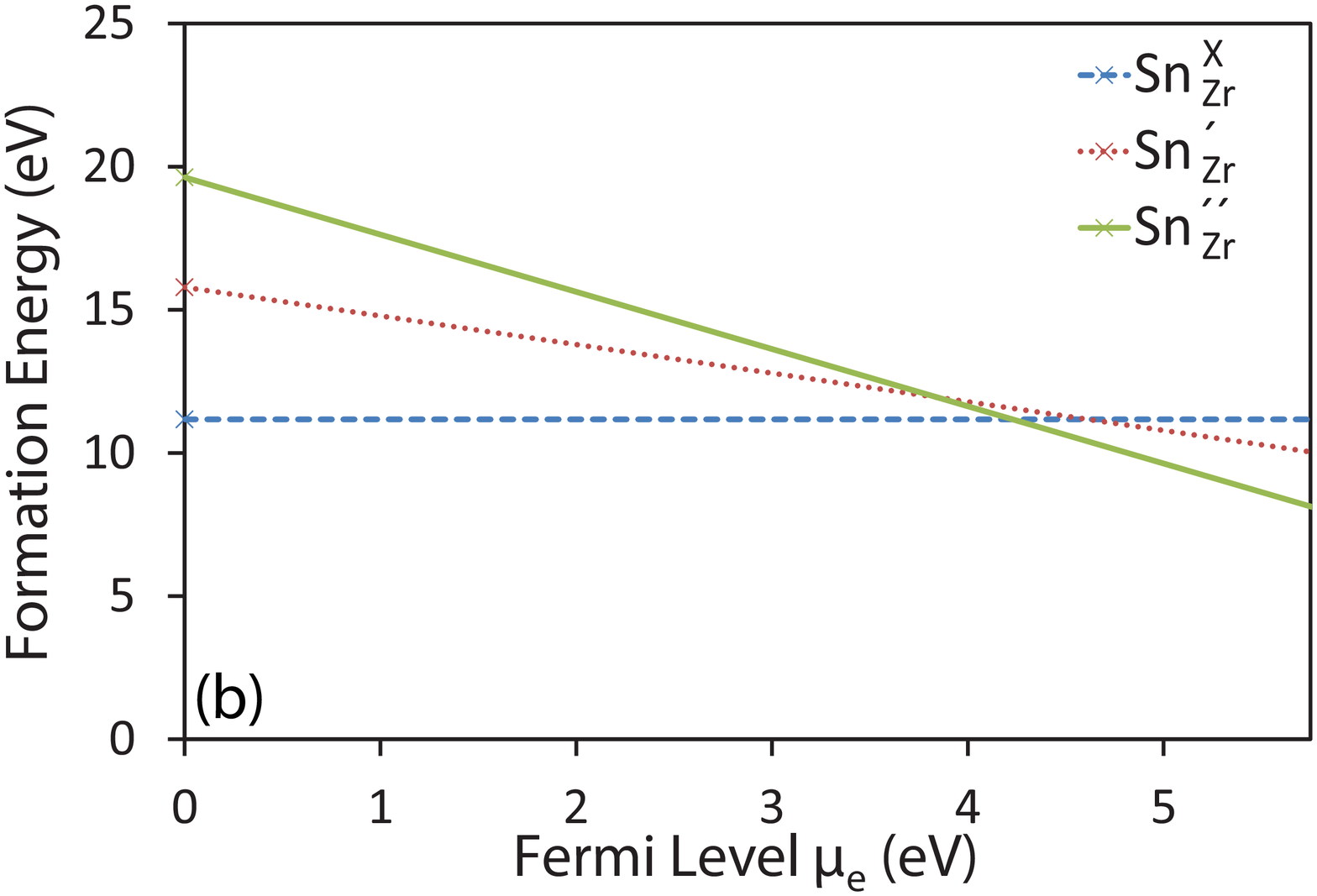}\hfill{}

\caption{Formation energies of intrinsic defects (a) and tin substitutional
defects (b) as calculated by Equation~\ref{eq:formation_intrinsic}
using DFT calculated energies from a reference state of Zr\textsubscript{(s)}
and plotted as a function of the Fermi level ($\mu_{e}$), from the
VBM across the experimental band gap of 5.75~eV~\cite{French1994}.
\label{fig:Formation-energies}}

\end{figure}

The defect formation energies for single isolated defects as a function
of $\mu_{e}$ calculated using Equation~\ref{eq:formation_intrinsic}
are shown in Figure~\ref{fig:Formation-energies}. Oxygen defects
exist almost entirely in the fully charged or zero charged states,
implying that the singly charged state is not thermodynamically favourable
for either a vacancy or interstitial.

Zirconium vacancies exist in the -4 oxidation state across the majority
of the band gap, with a transition through all charge states to 0
observed close to the VBM. The zirconium interstitial is also observed
in the fully charged +4 state across the majority of the band gap,
transitioning to +2 and then 0 close to the conduction band minimum
(CBM). These observations agree well with previous simulation work
investigating tetragonal ZrO\textsubscript{2}~\cite{Youssef2012,Eichler2001,Ganduglia-Pirovano2007a}.

Tin interstitial defects are predicted to have a very high formation
energy when compared to other possible defects and so can be assumed
to exhibit a negligible concentration. The formation energies for
the substitutional $Sn_{Zr}$ defect are shown in Figure~\ref{fig:Formation-energies}.
The $\mbox{\ce{S\ensuremath{n_{Zr}^{\times}}}}$ defect is favoured
across the majority of the band gap, with a transition to $\mbox{\ce{S\ensuremath{n_{Zr}^{''}}}}$
occurring close to the CBM. The $\mbox{\ce{S\ensuremath{n_{Zr}^{\prime}}}}$
defect is not observed at any point in the band gap, this is expected
as it would otherwise require tin to be in the +3 charge state which
would result in an energetically unfavourable unpaired s-orbital electron.

\subsection{Brouwer diagram}

\begin{figure}[H]
\hfill{}\includegraphics[width=0.5\textwidth]{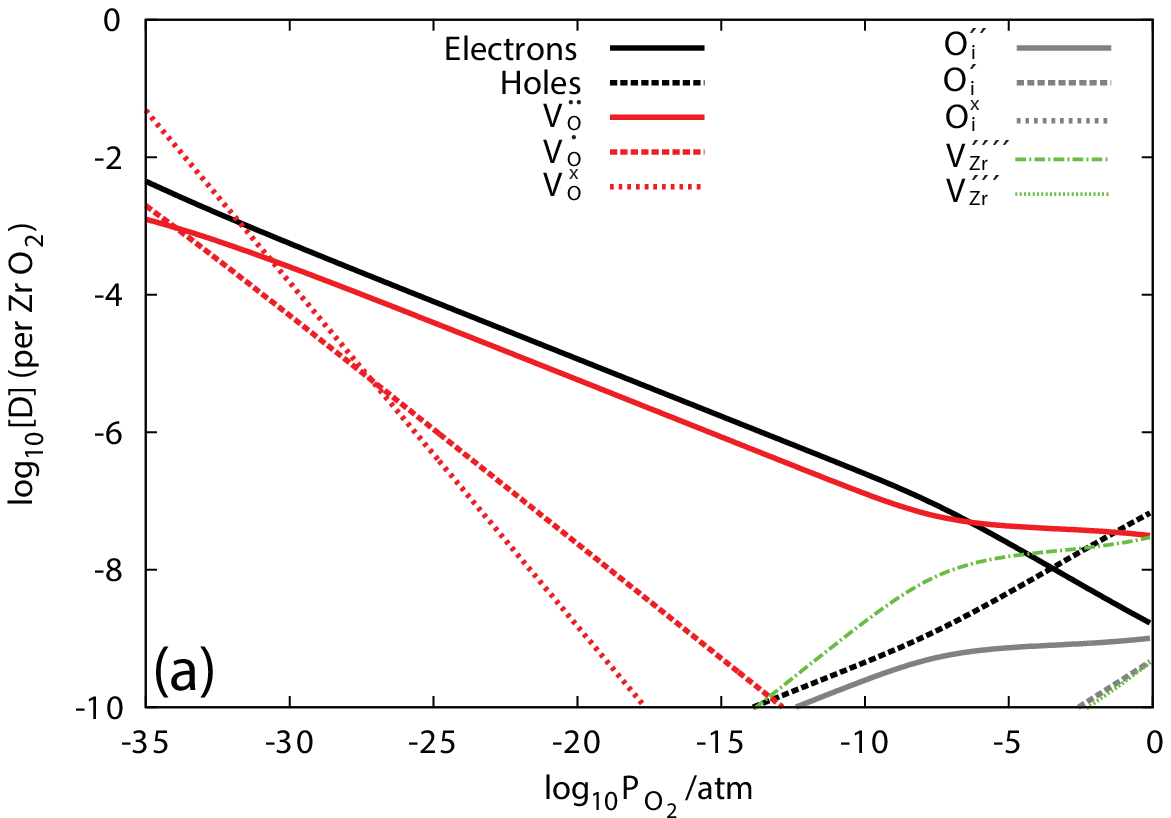}\hfill{}\includegraphics[width=0.5\textwidth]{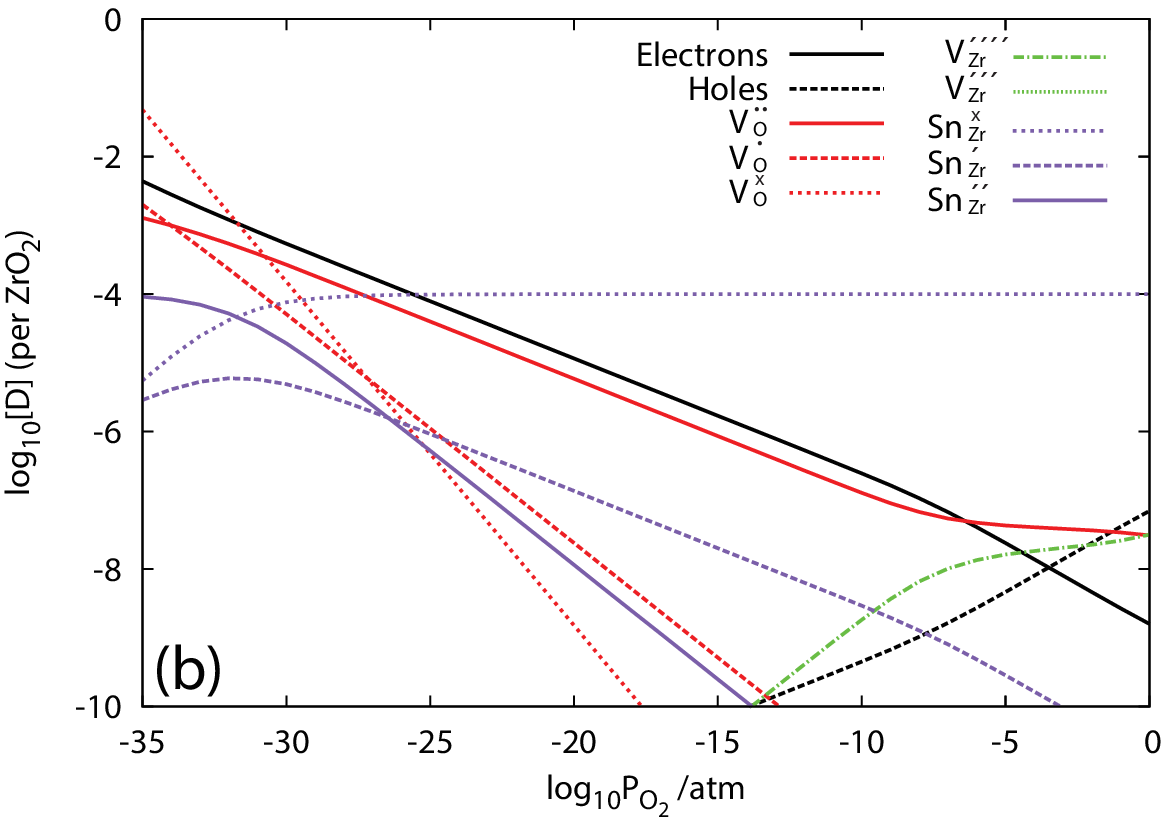}\hfill{}

\caption{Brouwer diagram showing the concentrations of point defects in tetragonal
ZrO\textsubscript{2} as a function of oxygen partial pressure at
1500K. (a) is the intrinsic system, (b) contains tin at a concentration
of $1\times10^{-4}$~at.\%. \label{fig:brouwer_diagrams}}
\end{figure}

Following the methodology discussed in Section~\ref{sub:Brouwer-Diagram},
Brouwer diagrams were constructed for intrinsic tetragonal ZrO\textsubscript{2}
and for tin doped ZrO\textsubscript{2}, and are shown in Figure~\ref{fig:brouwer_diagrams}.
As discussed in Section~\ref{sub:Oxide-phase-stability}, tetragonal
ZrO\textsubscript{2} is stabilised in the undoped oxide layer by
a combination of grain size and compressive stress. In this work,
stress was not applied and so the Brouwer diagrams were plotted at
1500~K, the approximate temperature at which tetragonal phase is
stable under standard conditions, rather than at a normal reactor
operation temperature of around 600~K. In this regard we follow the
approach used in previous DFT studies on the tetragonal system~\cite{Otgonbaatar2014,Youssef2012}.

In the intrinsic case (Figure~\ref{fig:brouwer_diagrams}a) at very
low oxygen partial pressures the uncharged oxygen vacancy appears.
However, across the majority of the oxygen partial pressures considered
the dominant structural defect is the fully charged oxygen vacancy,
charge compensated by electrons. Only at the highest oxygen partial
pressures does the fully charged zirconium vacancy, charge compensated
by holes, begin to appear.

The diagram that includes substitutional tin (Figure~\ref{fig:brouwer_diagrams}b)
indicates that $\mbox{\ce{S\ensuremath{n_{Zr}^{\times}}}}$ is dominant
across the majority of oxygen partial pressures considered, with a
transition to $\mbox{\ce{S\ensuremath{n_{Zr}^{''}}}}$ charge compensated
by $\mbox{\ce{\ensuremath{V_{O}^{\bullet\bullet}}}}$ occurring at
partial pressures below $10^{-31}$~atm.

\subsection{Tin interaction with other alloying elements}

Tin has been shown to have little effect on the corrosion resistance
and no measurable effect on HPUF in Zr-Sn binary alloys~\cite{Berry1961}.
As previously discussed, the tin Brouwer diagrams (Figure~\ref{fig:brouwer_diagrams})
suggest that tin exists as $\mbox{\ce{S\ensuremath{n_{Zr}^{\times}}}}$
until extremely low oxygen partial pressures. Given that $\mbox{\ce{S\ensuremath{n^{4+}}}}$
and $\mbox{\ce{Z\ensuremath{r^{4+}}}}$ occupy the same lattice site,
have similar ionic radii and exist in the same oxidation state it
is perhaps unsurprising that the $\mbox{\ce{S\ensuremath{n_{Zr}^{\times}}}}$
defect has little measurable effect on the chemistry of the system.
However, Sn-Zr binary alloys are not used in any reactor applications
and so we must consider the effect of further alloying additions.

Niobium containing alloys have been developed with excellent corrosion
and HPUF resistance and recent experimental work has suggested that
the removal of tin, included for its positive effect on the mechanical
properties of the alloys, improves the corrosion resistance still
further \cite{Wei2013}. Given that tin in isolation has no effect
on the corrosion and HPUF properties, it seems plausible that there
is an interaction between tin and niobium in the alloys.

Niobium is generally assumed to exist in the $\mbox{\ce{N\ensuremath{b^{5+}}}}$
state within the oxide layer. If we assume the 5+ state, this results
in a predominant $\mbox{\ce{N\ensuremath{b_{Zr}^{\bullet}}}}$ defect,
which agrees with DFT simulations by Otgonbaatar~\textit{et~al.}~\cite{Otgonbaatar2013}.
This would lead to the injection of positive charge into the oxide
layer. Nevertheless, this assumption has been challenged recently
through XANES work which instead suggests the charge state is between
2+ and 4+~\cite{Froideval2008,Couet2014}. Recent theoretical and
experimental work by Couet~\textit{et~al.} \cite{CouetSubmitted}has
suggested another possible source of positive charge close to the
metal-oxide interface due to a space charge effect caused by the non-equilibrium
distribution of electrons in the insulating oxide layer.

\begin{figure}
\hfill{}\includegraphics[width=0.5\textwidth]{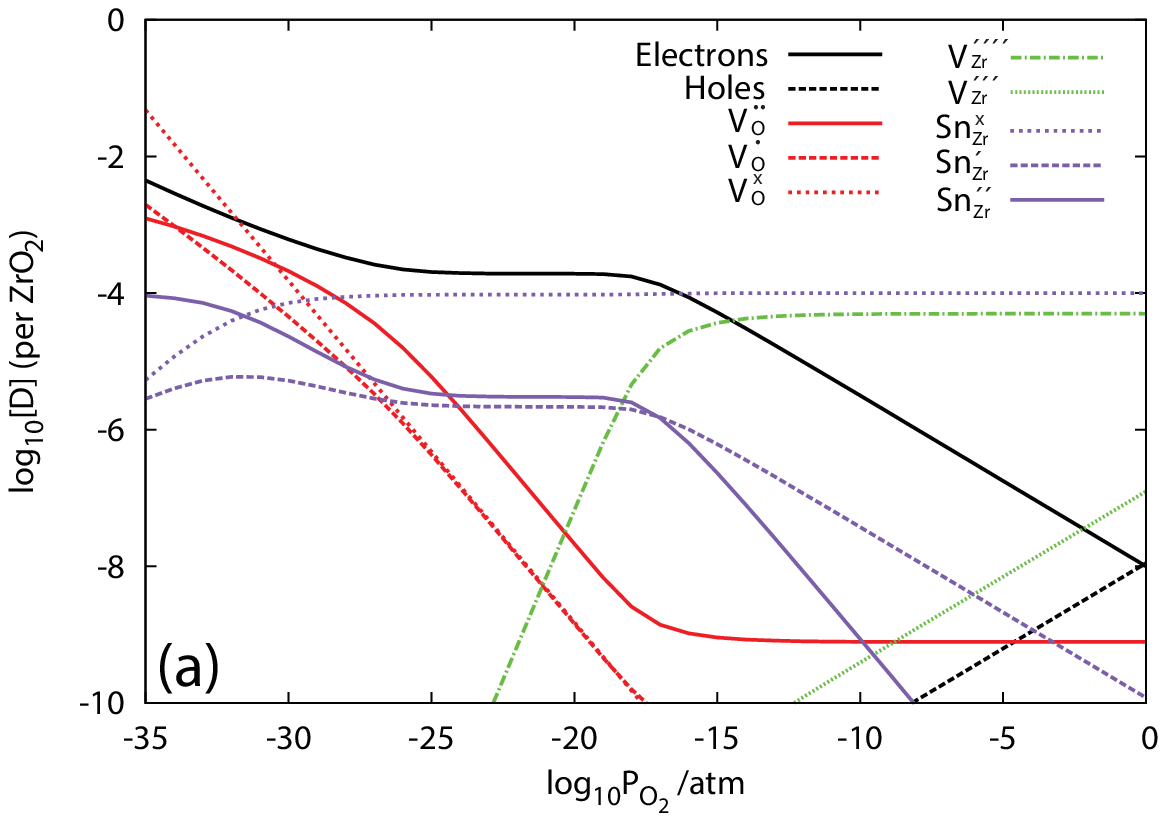}\hfill{}\includegraphics[width=0.5\textwidth]{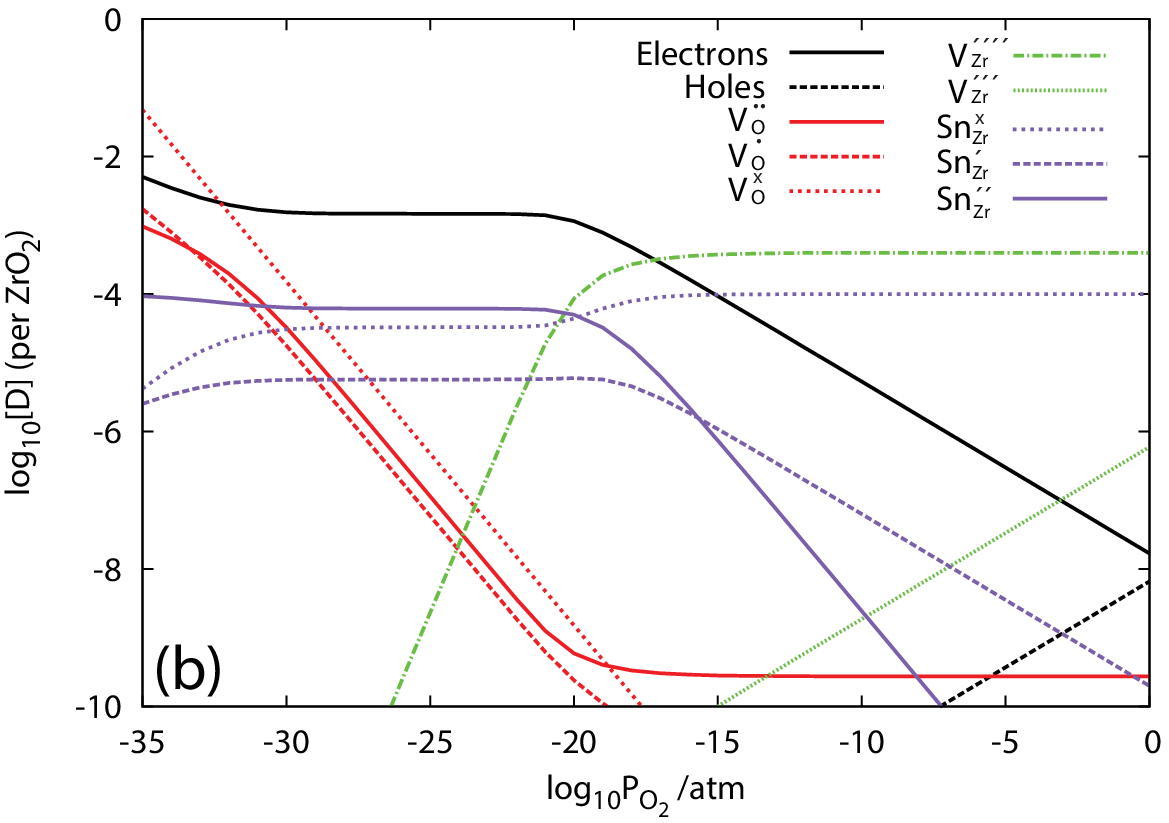}\hfill{}

\caption{Brouwer diagrams containing tin at a concentration of $1\times10^{-4}$~at.\%
and an additional defect of charge +1 ($q=1$) inserted into the calculation
via Equation~\ref{eq:charge_neutrality}; (a) $c=5\times10^{-4}$,
(b) $c=2\times10^{-3}$. \label{fig:Brouwer-diagrams-charged}}
\end{figure}

In order to consider the effect of the positive charge, an additional
defect was added to the first term of Equation~\ref{eq:charge_neutrality}.
This defect was given a charge of +1, and a Brouwer diagram was plotted
for a range of concentrations. The inclusion of an additional positive
defect at a concentration of $5\times10^{-4}$ (Figure~\ref{fig:Brouwer-diagrams-charged}a)
causes the concentration of $\mbox{\ce{\ensuremath{V_{Zr}^{''''}}}}$
to dramatically increase and also lowers the concentration of $\mbox{\ce{\ensuremath{V_{O}^{\bullet\bullet}}}}$
and $\mbox{\ce{\ensuremath{V_{O}^{\bullet}}}}$ (the concentration
of $\mbox{\ce{\ensuremath{V_{O}^{\times}}}}$ is unchanged since it
is not charged and therefore unaffected by charge difference in the
system). Importantly, although the concentration of $\mbox{\ce{S\ensuremath{n_{Zr}^{''}}}}$
is increased, $\mbox{\ce{S\ensuremath{n_{Zr}^{\times}}}}$ remains
the dominant defect at partial pressures above $10^{-31}$~atm.

\begin{figure}[H]
\hfill{}\includegraphics[width=0.5\textwidth]{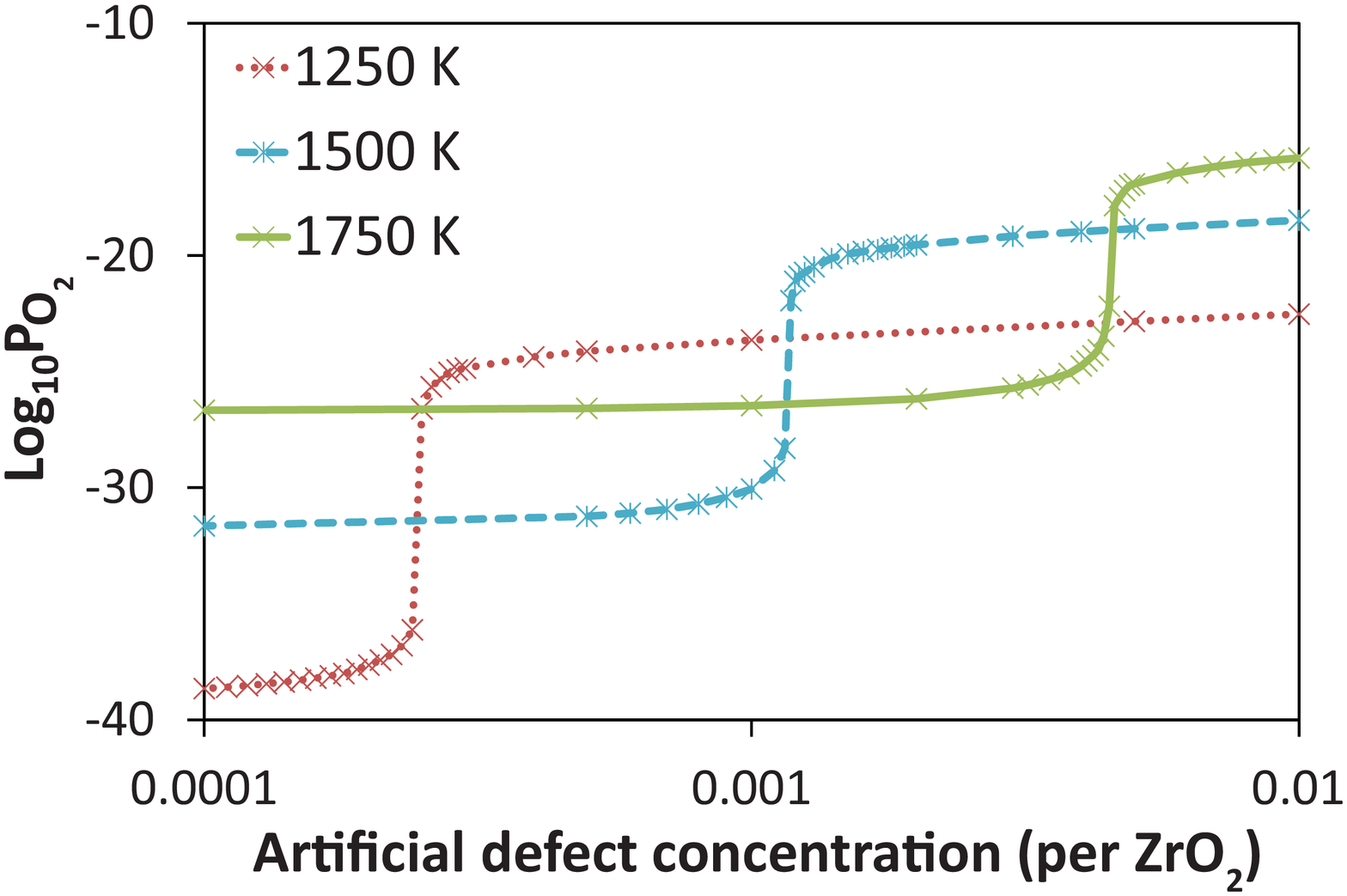}\hfill{}

\caption{A chart showing the oxygen partial pressure at which the $Sn_{Zr}^{\times}$/$Sn_{Zr}^{''}$
transition occurs as a function of the applied additional charge;
1250~K at a concentration of $(2.4-2.5)\times10^{-4}$, 1500~K at
$(1.15-1.18)\times10^{-3}$, 1750~K at $(4.5-4.6)\times10^{-3}$
\label{fig:sn_crossover}}
\end{figure}

Increasing the concentration of applied charge further (Figure~\ref{fig:Brouwer-diagrams-charged}b)
causes $\mbox{\ce{S\ensuremath{n_{Zr}^{''}}}}$ to become the dominant
tin defect up to a partial pressure of $10^{-20}$~atm. This increase
of around 10 orders of magnitude happens at a critical charge concentration
and as demonstrated by Figure~\ref{fig:sn_crossover} even large
changes in charge above or below this critical value have little effect
on the crossover point. The critical charge at which transition occurs
appears to be strongly dependent on temperature, with an increase
of 500~K causing the value to increase by an order of magnitude,
suggesting that the $\mbox{\ce{S\ensuremath{n_{Zr}^{\times}}}}$ defect
is more favourable at higher temperatures. The existence of a critical
value has previously been predicted in sapphire by Lagerlöf~and~Grimes~\cite{Lagerlof1998},
who observed that a critical doping level caused the concentration
of oxygen vacancies to change by 10-20 orders of magnitude. Interestingly,
as the temperature of the Brouwer diagram calculation was increased
the change in partial pressure observed at the critical doping level
reduced.

\subsection{Implications for corrosion resistance and HPUF}

It is generally assumed that niobium improves corrosion resistance
by suppressing the formation of oxygen vacancies in the bulk oxide,
thereby reducing the oxygen ion conductivity. This is achieved due
to the positive charge introduced by the dominant $\mbox{\ce{N\ensuremath{b_{Zr}^{\bullet}}}}$
defect, which suppresses $\mbox{\ce{V\ensuremath{{}_{O}^{\bullet\bullet}}}}$
and $\mbox{\ce{V\ensuremath{{}_{O}^{\bullet}}}}$ since they are also
positively charged. However, as demonstrated in Figures~\ref{fig:Brouwer-diagrams-charged}~and~\ref{fig:sn_crossover},
the presence of a concentration of positive charge in the system causes
the $\mbox{\ce{S\ensuremath{n_{Zr}^{\times}}}}$ defect to become
$\mbox{\ce{S\ensuremath{n_{Zr}^{''}}}}$. Considering the oxide layer
on a Zr-Sn-Nb type alloy, this tin behaviour would negate the oxygen
vacancy suppression delivered by niobium by charge compensating $\mbox{\ce{N\ensuremath{b_{Zr}^{\bullet}}}}$
defects, suggesting a possible mechanism by which tin diminishes the
corrosion resistance of Zr-Sn-Nb alloys.

As mentioned in the introduction, oxygen vacancies can have the effect
of stabilising the tetragonal phase. Thus, a further interesting implication
is that the increased $\mbox{\ce{V\ensuremath{{}_{O}^{\bullet\bullet}}}}$
concentration in Sn-containing Nb-Zr alloys could result in the increased
tetragonal phase fraction observed in XRD work by Wei~\textit{et~al.}~\cite{Wei2013}.
As corrosion progresses, the metal-oxide interface will move further
away from the $\mbox{\ce{S\ensuremath{n_{Zr}^{''}}}}$ substitutional
defects interface due to the thickening of the oxide layer, which
will have the effect of increasing the oxygen partial pressure in
the oxide surrounding the defects. As demonstrated by Figures~\ref{fig:Brouwer-diagrams-charged}~and~\ref{fig:sn_crossover},
there is a critical partial pressure at which the $\mbox{\ce{S\ensuremath{n_{Zr}^{''}}}}$
defect will reduce to $\mbox{\ce{S\ensuremath{n_{Zr}^{\times}}}}$.
As this reduction occurs, the increased $\mbox{\ce{V\ensuremath{{}_{O}^{\bullet\bullet}}}}$
concentration associated with the presence of $\mbox{\ce{S\ensuremath{n_{Zr}^{''}}}}$
is no longer expected. Thus, the stabilisation mechanism causing the
increased tetragonal phase volume will no longer exist and a transformation
to monoclinic phase will occur. The tetragonal to monoclinic phase
transformation is associated with a volume increase of approximately
4\%~\cite{Heuer1985} and so cracking of the oxide layer is expected.
This process could lead to an earlier onset of transition than would
otherwise occur, a conclusion that agrees closely with experimental
observations that the reduction of Sn content delays the first transition
significantly~\cite{Wei2013}.

\section{Conclusions}

DFT simulations were used to investigate the defect properties of
intrinsic and Sn-doped tetragonal ZrO\textsubscript{2}. The intrinsic
case agreed well with previous DFT work performed by Youssef~\textit{et~al.}~\cite{Youssef2012}.
Tin is predicted to exhibit a 4+ charge state and exist almost entirely
in uncharged $\mbox{\ce{S\ensuremath{n_{Zr}^{\times}}}}$, transitions
to a 2+ charge state and thus $\mbox{\ce{S\ensuremath{n_{Zr}^{''}}}}$
only at oxygen partial pressures below $10^{-31}$~atm.

An additional positively charged defect was included in the Brouwer
diagram calculations to account for the effect of positively charged
defect species, in particular $\mbox{\ce{N\ensuremath{b_{Zr}^{\bullet}}}}$.
It is predicted that, up to a critical concentration, the additional
defect has little effect on the dominance of $\mbox{\ce{S\ensuremath{n_{Zr}^{\times}}}}$
defect, with the only observed effect being the suppression of oxygen
vacancies and increased concentration of zirconium vacancies, as would
be expected in a Nb-containing system. Above a temperature dependent
critical concentration, however, the additional positive charged defect
promotes the transition from $\mbox{\ce{S\ensuremath{n_{Zr}^{\times}}}}$
to $\mbox{\ce{S\ensuremath{n_{Zr}^{''}}}}$ to occur at an oxygen
partial pressure many orders of magnitude higher. The concentration
at which this change occurs and the magnitude of the change in partial
pressure has a strong temperature dependence; a higher charge concentration
was required at high temperatures but resulting in a smaller shift
in partial pressure.

It is suggested that this change in tin oxidation state will inhibit
the oxygen vacancy suppression, which would otherwise be caused by
the additional positive charge, thereby accounting for the increased
corrosion rate and shorter time to transition observed in Zr-Sn-Nb
type alloys when compared to similar alloys containing niobium but
no tin. With no niobium content, we expect tin to maintain an overall
neutral charge ($\mbox{\ce{S\ensuremath{n_{Zr}^{\times}}}}$) and
therefore have little effect on the corrosion performance or time
to transition. This provides a model framework by which we can interpret
experimental work on Zn-Sn binary alloys~\cite{Berry1961}.

\section*{Acknowledgements}

The authors would like to thanks Rolls-Royce for the funding of this
work as part of the Westinghouse led MUZIC-2 research programme and
for the computational resources provided by the Imperial College High
Performance Computing Centre.

\end{doublespace}

\bibliographystyle{unsrt}
\bibliography{sn_in_t-zro2}

\end{document}